\documentclass[12pt,a4paper]{article}
\usepackage{subeqn}
\usepackage{amsfonts}
\usepackage{defns}
\usepackage[colour,hyperref]{ajr}

\iffalse\newcommand{\afivepage}{\newpage}
\else\newcommand{\afivepage}{}
\fi

\renewcommand{\marginpar}[1]{}

\newcommand{\dn}{{\mathfrak D}}%{\mbox{Dn}}
\newcommand{\vel}{\vec q}
\newcommand{\re}{{\cal R}}
\newcommand{\rr}{{\mathfrak R}}
\newcommand{\repe}{\alpha}
\newcommand{\pe}{{\cal P}}
\newcommand{\eu}{\vec u}
\newcommand{\ep}{\vec p}
\newcommand{\eb}{\vec b}
\newcommand{\tp}{\theta}
\newcommand{\tht}{\vartheta}
\newcommand{\red}{\textsc{reduce}}
\newcommand{\omrr}{\left(1-r^2\right)}
\newcommand{\dauq}{\hspace*{-1em}}

\begin{document}
\title{Shear dispersion along circular pipes is affected by bends, but
the torsion of the pipe is negligible} \author{A.~J.~Roberts\thanks{Dept
of Maths \& Computing, University of Southern Queensland, Toowoomba,
Queensland 4350, \textsc{Australia}.
\protect\url{mailto:aroberts@usq.edu.au}}}

\date{September 9, 2003}

\maketitle

\begin{abstract}
The flow of a viscous fluid along a curving pipe of fixed radius is
driven by a pressure gradient.  For a generally curving pipe it is the
fluid flux which is constant along the pipe and so I correct fluid flow
solutions of Dean~(1928) and Topakoglu~(1967) which assume constant
pressure gradient.  When the pipe is straight, the fluid adopts the
parabolic velocity profile of Poiseuille flow; the spread of any
contaminant along the pipe is then described by the shear dispersion
model of Taylor~(1954)\nocite{Taylor54} and its refinements by Mercer,
Watt~\etal~(1994,1996).\nocite{Mercer94a,Watt94c}  However, two conflicting
effects occur in a generally curving pipe: viscosity skews the velocity
profile which enhances the shear dispersion; whereas in faster flow
centrifugal effects establish secondary flows that reduce the shear
dispersion.  The two opposing effects cancel at a Reynolds number of
about~$15$.  Interestingly, the torsion of the pipe seems to have very
little effect upon the flow or the dispersion, the curvature is by far
the dominant influence.  Lastly, curvature and torsion in the fluid
flow significantly enhance the upstream tails of concentration profiles
in qualitative agreement with observations of dispersion in river flow.
\end{abstract}

\tableofcontents

\section{Introduction}
\label{sec:intro}

\begin{figure}[tbp]
	\centering
    \begin{tabular}{cc}
	\includegraphics[width=0.55\textwidth]{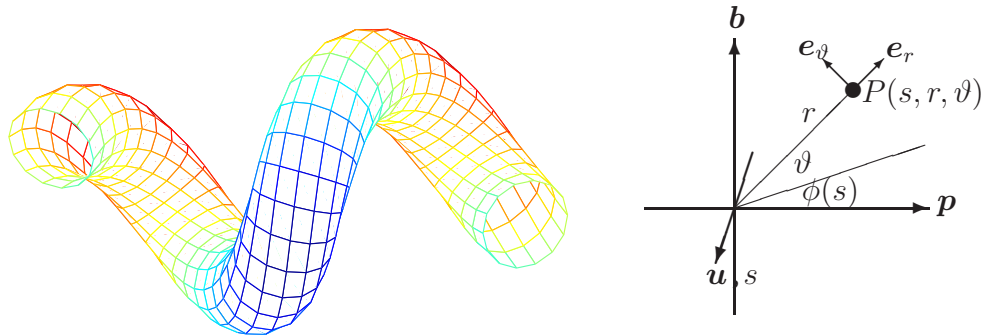}
    &
    {\tt    \setlength{\unitlength}{0.15ex}
	\begin{picture}(178,169)
	\thinlines    \put(116,114){$P(s,r,\tht)$}
	              \put(85,137){$\vec e_\tht$}
	              \put(128,136){$\vec e_r$}
	              \put(112,120){\vector(-1,1){15}}
	              \put(112,120){\vector(1,1){15}}
	              \put(83,79){$\tht$}
	              \put(54,62){\line(3,1){93}}
	              \put(54,62){\line(1,1){58}}
	              \put(87,104){$r$}
	              \put(112,120){\circle*{8}}
	              \put(87,66){$\phi(s)$}
	\thicklines   \put(51,150){$\vec b$}
	              \put(153,61){$\vec p$}
	              \put(40,25){$\vec u$,$s$}
	              \put(63,89){\vector(-1,-3){18}}
	              \put(54,10){\vector(0,1){135}}
	              \put(10,62){\vector(1,0){139}}
	\end{picture}}
    \end{tabular}
	\caption{perspective figure of a helical pipe as an example of the
	curving pipes containing fluid flow and contaminant dispersion that
	is modelled herein.  On the right is a schematic diagram of the
	orthogonal curvilinear coordinate system local to the curving
	centre line of the pipe: $\vec u$ points along the centre line and
	$s$~measures axial distance.}
	\label{fig:tubesurf}
\end{figure}

Consider the dispersion of a contaminant, with
diffusivity~$\kappa_c$, in the steady laminar flow, velocity
field~$\vel$, of a Newtonian fluid of density~$\rho$ and kinematic
viscosity~$\nu$ in an arbitrarily curving pipe of radius~$a$ such as
the helical pipe shown in Figure~\ref{fig:tubesurf}.  The flow is
pumped by an overall pressure drop which maintains a fixed fluid flux
along the pipe; that is, a constant mean velocity~$U$ is maintained
irrespective of curvature.  Non-dimensionalise quantities with respect
to the pipe radius~$a$, the cross-pipe diffusion time~$a^2/\kappa_c$,
and the reference pressure~$\rho\nu U/a$.  The Navier-Stokes and
continuity equations for the steady, incompressible fluid flow then
become
\begin{equation}
	\re\vel\cdot\grad\vel=-\grad p+\nabla^2\vel
	\quad\mbox{and}\quad
	\divv\vel=0\,,
	\label{eq:ns}
\end{equation}
where $\re=aU/\nu$ is the Reynolds number.  The contaminant evolves 
according to the non-dimensional advection-diffusion equation
\begin{equation}
	\D tc+\pe\vel\cdot\grad c=\nabla^2c\,,
	\label{eq:ad}
\end{equation}
where $\pe=aU/\kappa_c$ is the Peclet number.  In typical liquids the
Peclet number is much larger than the Reynolds number as their ratio,
the Schmidt (or Prandtl) number $\sigma=\pe/\re=\nu/\kappa_c$, is
normally large: approximately~$10^3$ for the diffusion of material in
liquids \cite[p1119, e.g.]{Ruthven71}; although only about~$8$ for the
diffusion of heat \cite[p597]{Batchelor67};\footnote{The Prandtl number
of water is~$13.4$ at~$0^\circ$, $9.5$ at~$10^\circ$, $8.1$
at~$15^\circ$, $7.1$ at~$20^\circ$, $5.5$ at~$30^\circ$, $4.3$
at~$40^\circ$, $3.0$ at~$60^\circ$, and $2.2$ at~$80^\circ$.  Whereas
the Prandtl number of air is~$0.71$\,.} whereas in typical gases the
Schmidt number is roughly~$1$ and so the Peclet and Reynolds numbers
are comparable.  The analysis is interpreted with these two cases in
mind of the Schmidt number being either $\Ord{1000}$ or~$\Ord{1}$.

Here we analyse the flow and dispersion in an arbitrarily curving
circular pipe.  Most analysis of dispersion assumes a curved pipe is
toroidal \cite[e.g.]{Ruthven71, Nunge72, Johnson86, Daskopoulos89} and
most experiments are performed in helical pipes \cite[e.g.]{Trivedi75}
(see further discussion by Berger \etal~\cite{Berger83}).  Neglecting molecular
diffusivity, dispersion in toroidal flow has been characterised from
analytic formulae by Ruthven~\cite{Ruthven71} and numerical solutions by
McConalogue~\cite{McConalogue70} using the residence time of different
streamlines.  Interestingly, using similar ideas Jones~\cite{Jones94}
deduced a regime of anomalous dispersion in a twisted but piecewise
toroidal pipe.  The fluid flow in helical pipes has been the subject of
recent analysis \cite[e.g.]{Germano82, Tuttle90, Liu93, Zabielski98a},
whereas the flow in arbitrarily curving and twisting pipes has received
little attention although Pedley~\cite{Pedley80} accounted for leading
order effects of variable curvature but ignored torsion, and Gammack \&
Hydon~\cite{Gammack01} investigate flow in pipes with exponentially
varying curvature and torsion.  Here I take these analyses further by
simultaneously determining the fluid flow and the dispersion in
arbitrarily curving circular pipes.  The analysis is restricted to
parameter regimes where the fluid flow is laminar---because of the
induced secondary circulation, laminar flow is stable to higher
Reynolds numbers in a curving pipe \cite{Trivedi75, Berger83}.  The
results thus will also be important in the flow and dispersion in
microfluidic channels~\cite[e.g.]{Ghosal02}.

In \S\ref{SScoord} we establish an orthogonal coordinate
system based upon the arbitrarily curving geometry of the circular pipe,
although requiring that the curvature of the centre line varies
smoothly.  Let the centre line of the pipe be described by~$\vec R(s)$
where $s$~measures arclength along the centre line.  Then a useful set
of vectors for space in the vicinity of the pipe are the unit tangent
$\eu(s)=\vec R'$ of the centre curve, unit normal~$\ep(s)$ and the unit
binormal~$\eb(s)$ (see Figure~\ref{fig:tubesurf}).  These vectors and
the curvature~$\kappa(s)$ and torsion~$\tau(s)$ of the pipe are all
connected by the Frenet formulae \cite[\S8.7,e.g.]{Kreyszig8}
\begin{equation}
	\eu' = \kappa\ep\,,\quad
	\ep' = -\kappa\eu+\tau\eb\,,\quad
	\label{eq:frenet} 
	\eb' = -\tau\ep\,.
\end{equation}
Throughout this article I use a dash to denote~$\partial/\partial s$.
In a thin pipe the non-dimensional velocity is approximately 
Poiseuille flow, $u\approx 2(1-r^2)$.
However, there are corrections of~$\Ord{\kappa}$ due to the curvature 
which are determined by solving the Navier-Stokes 
equations~(\ref{eq:ns}) for the fluid flow, see \S\ref{Sns} where 
low order expressions agree with the careful analysis of 
the flow in helical pipes by Tuttle~\cite{Tuttle90}.

Centre manifold theory provides a rationale to form low-dimensional
models of dynamics as I elaborated in the overview~\cite{Roberts03b}.
Here we model the long-time evolution of the large scale dispersion of
contaminant along the pipe.  Models of ``long-waves'' or ``slowly
varying in space'' dynamics are justified in the centre manifold
approach by requiring resolved longitudinal spatial structures to have
small wavenumber~\cite{Roberts88a}.  With this proviso, diffusion acts
relatively quickly across the pipe to cause the contaminant
concentration to be approximately constant in any cross-section: to
leading order $c\approx C(s,t)$ where $C=\bar c$ is an average over a
pipe cross-section.  The centre manifold analysis then systematically
accounts for how variations along the pipe are affected by the varying
velocity profile to disperse the contaminant.  We thus deduce in
\S\ref{Sdisp}, as a generalisation of Taylor's model
\cite{Taylor53, Taylor54}, the advection-diffusion model
\begin{equation}
	\D tC\approx -\pe\D sC+\D s{\ }\left(D\D sC\right)\,,
	\label{eq:tay}
\end{equation}
where for the case of a pipe of circular cross-section the effective 
diffusivity
\begin{eqnarray}
	D&\!\!\!=\!\!\!&\left(1+\frac{\kappa^2}{4}\right)
	+\frac{\pe^2}{48}\left[ 1 +\kappa^2\left( \frac{863}{120} 
	-\frac{7267\re^2}{241920} +\frac{599\re^4}{48384000} 
	-\frac{2569\sigma^2\re^4}{68428800} \right) \right]
	\nonumber\\&&
	{}+\Ord{\kappa^4,\delta}\,.
	\label{eq:difad}
\end{eqnarray}
That is, the shear dispersion in a straight pipe, $D=\pe^2/48$, is in 
a curved pipe modified by a factor approximately
\begin{displaymath}
	1+\kappa^2\left[ 7.2-3.0\left(\frac{\re}{10}\right)^2
	+(0.12-0.38\sigma^2)\left(\frac{\re}{10}\right)^4 
	\right]\,,
\end{displaymath}
As found by others, secondary circulation caused by fluid
inertia depresses the effective dispersion, by about
$\kappa^2[3.0(\re/10)^2+0.38\sigma^2(\re/10)^4]$\,, but only for
Reynolds numbers sufficiently large.  In slow viscous flow, pipe
curvature actually enhances the effective dispersion by
about~$7.2\kappa^2$---an effect that has apparently often been
neglected \cite[p317]{Trivedi75} in experimental determination of
dispersion coefficients.
\begin{itemize}
\item For dispersion in gas flow, with $\Ord{1}$~Schmidt
number~$\sigma$, the dispersion is depressed by secondary circulation
only if $\re$~is greater than about~$15$ as otherwise the viscous
enhancement is stronger.  Remarkably, if the Schmidt number~$\sigma$ is
small, less than about~$0.5$, inertial effects in the fluid flow
enhance the effective dispersion for $\re$~greater than about~$50$.

\item For dispersion in liquids, with say $\Ord{1000}$ Schmidt
number~$\sigma$, the term in $\kappa^2\re^4\sigma^2$ dominates the
other terms for Reynolds number greater than about~$0.5$.  Hence, in
liquids and due to secondary circulations due to inertia, I reaffirm
the reduction in effective dispersion.
\end{itemize}
Since the Dean number\footnote{As discussed by 
Berger \etal~\cite[\S2.1.1.2]{Berger83}, there are various and conflicting 
definitions of the Dean number: Berger \etal\ recommended the use of 
$\dn=2\sqrt\kappa\re$ which I have adopted here.  This Dean number 
could be viewed as~$\sqrt\kappa\re$ for a Reynolds number based upon 
the pipe diameter rather than the radius that I have used.  } 
$\dn=2\sqrt\kappa\re$ this last is the shear dispersion correction 
verified by Nunge~\cite{Nunge72} and Johnson~\cite{Johnson86} as being 
significant for $\dn ^2\sigma$ greater than about~$100$.  A limitation 
of the expression~(\ref{eq:difad}) is that it predicts a physically 
unrealisable negative effective diffusivity for large enough Reynolds 
number.  In most cases, Schmidt number~$\sigma$ larger than~$1$, the 
term in $\kappa^2\sigma^2\re^2$ dominates the correction.  Hence to 
maintain a positive diffusion coefficient the Dean number 
$\dn<25/\sqrt\sigma$ or equivalently $\sigma\dn^2<650$\,.  The 
expression~(\ref{eq:difad}) is a low order approximation to the correct 
curve, describing the downwards curving shape on the left side of 
Figure~3 of Johnson~\cite{Johnson86}, but needing the higher order 
corrections described in \S\ref{SShigh} to describe the 
dispersion coefficient at higher Dean number.
% Some of the 
% spread in Johnson \& Kamm's Monte Carlo simulations may be due to the 
% varying relative importance in the terms as the Schmidt number varies.

For lower Reynolds number the qualitative deductions above vary from
those of Nunge~\cite{Nunge72} because their dispersion coefficient is
different, see their equation~(76).  In particular, I predict that
shear dispersion is frequently enhanced for gases, the reverse
conclusion to that of Erdogan~\cite{Erdogan67} and later
Nunge~\cite[pp.363,375]{Nunge72}.  I argue that the differences occur
because all previous work, based upon the fluid flow solutions of
Dean~\cite{Dean28, Dean27} and Topakoglu~\cite{Topakoglu67}, have
assumed that the pressure \emph{gradient} is fixed in the expansion in
curvature~$\kappa$---an adequate assumption for flow in a torus or
helix where the curvature and the torsion are constant.  But in a pipe
of generally varying curvature and torsion, as developed here, it is
the mean fluid flux which is fixed along the pipe, not the pressure
gradient.\footnote{Even Gammack \& Hydon \cite[p363]{Gammack01} appear
to fix the pressure gradient in their exponentially varying pipes by
requiring the second order pressure correction~$p_2\propto\sin\xi$\,,
where $\xi$~is their angular variable, and so their pressure correction
has zero mean.} Since, for a constant pressure gradient the fluid flux
varies with curvature and torsion---generally first decreasing with
increasing torsion then later increasing with torsion, see
Yamamoto~\cite{Yamamoto94} and its correction \cite{Yamamoto99}---it
follows that the mean pressure gradient~(\ref{eq:meanpgrad}) varies
along a generally curving pipe.  To check my computer algebra program
(listed in Appendix~\ref{Sca}) I temporarily fixed the pressure
gradient in a helical pipe and found the resulting dispersion
coefficient to be exactly equivalent to that given by
Nunge~\cite{Nunge72}, equation~(76), except that the one term in
$\re^2\sigma\kappa^{2}$ (my~$\kappa$ is their~$1/\lambda$) is zero in
my results---I conjecture theirs is in error in this term.  Because of
the requirement to fix the fluid flux I recommend the use
of~(\ref{eq:difad}) instead of the earlier published models of shear
dispersion.

The error of~$\Ord{\delta}$ in the shear dispersion coefficient given
by~(\ref{eq:difad}) encompasses modifications due to torsion~$\tau$ and
to variations in curvature~$\kappa$ along the pipe: the
parameter~$\delta$ corresponds to the parameter~$\eta$ in Gammack \&
Hydon's analysis of exponentially varying pipes, $\kappa\propto e^{\eta
s}$.  The torsion only affects the dispersion coefficient
at~$\Ord{\kappa^2\tau^2}$, as see \S\ref{Sdisp}, and so does not appear
in~(\ref{eq:difad}).  The effects of axial variations are reformulated
as memory of the effective dispersion coefficient some distance
upstream.  Such memory effects in shear dispersion in varying channels
were first recognised by Smith~\cite{Smith83b}.

Using computer algebra it is also straightforward to determine both
higher order corrections to the dispersion coefficient and high-order
terms in the advection-diffusion equation itself.  These terms may be
either used to refine the approximations, or to give good estimates of
the errors in a lower-order approximation.  Earlier work by Mercer \&
Roberts~\cite{Mercer94a} gave a sharp estimate for the limit of spatial
resolution in a straight circular pipe.

% \marginpar{More here ??}

\section{The fluid flow}
\label{Sns}

The first task is to find the laminar viscous fluid flow in the 
curving pipe.  The focus of the paper is the dispersion in the pipe by 
the flow, but there are enough interesting and relevant features in 
the fluid flow itself to be discussed briefly here---in particular 
this section confirms aspects of my analysis by reproducing many 
results of other authors about steady laminar flow in curved and 
twisted pipes.

We assume that the flow is steady as appropriate to flow driven by a 
constant pressure drop through a fixed pipe.  However, to maintain 
everywhere constant fluid flux, the mean pressure gradient, ${\bar 
p}'$, varies with the curvature of the pipe as given 
in~(\ref{eq:meanpgrad}).

\subsection{The orthogonal curvilinear coordinate system}
\label{SScoord}
Expressions for the flow are derived in an orthogonal curvilinear
coordinate system matched to the geometry of the circular pipe.  The
orthogonal coordinate system has been used by Germano~\cite{Germano82},
Kao~\cite{Kao87}, Liu~\cite{Liu93} and Yamamoto~\cite{Yamamoto94,
Yamamoto99} to investigate the structure of the fluid flow in helical
pipes up to Dean numbers of~$2,000$, and is well known in
hydromagnetodynamics.  One difference here is that we do not assume the
pipe is helical, instead we allow arbitrary variations in the curvature
and torsion of the pipe---the one important restriction is that the
curvature and torsion must vary only slowly along the tube.  Such slow
variations along the pipe were also assumed by Murata~\cite{Murata76}
in their analysis of the flow in tubes bent sinusoidally in a plane,
and Pedley~\cite{Pedley80} in a leading approximation to the effects of
curvature.  As shown schematically in Figure~\ref{fig:tubesurf},
positions in space are labeled by $(s,r,\tht)$ and have position vector
\begin{equation}
	\vec r=\vec R(s)+r\cos\tp\,\ep+r\sin\tp\,\eb\,,
	\quad\mbox{where}\quad\tp=\tht+\phi(s)
	\label{eq:posi}
\end{equation}
measures the angle from the plane of the normal~$\ep$ to the
point~$\vec r$; thus $\tp=0$ corresponds to the inside of the local
bend whereas $\tp=\pm\pi$ corresponds to the outside.  However, due to
torsion in the shape of the pipe the reference plane of the orthogonal
coordinate system must twist along the pipe by an amount~$\phi(s)$
where
\begin{displaymath}
	\frac{d\phi}{ds}=-\tau\,.
\end{displaymath}
The unit vectors and scale factors of this orthogonal 
coordinate system are then
\begin{equation}
	\begin{array}{ll}
		h_s=1-\kappa r\cos\tp\,, & \vec e_s=\eu\,,  \\
		h_r=1\,, & \vec e_r=\phantom{-}\cos\tp\,\ep+\sin\tp\,\eb\,,  \\
		h_\tht=r\,, & \vec e_\tht=-\sin\tp\,\ep+\cos\tp\,\eb\,.
	\end{array}
	\label{eq:he}
\end{equation}
Note that all expressions for fluid and concentration fields are
written in terms of~$\tp$, the angle relative to the local direction of
curvature of the bent pipe---because it is this angle that primarily
determines the shape of the local fields---but all equations are
written in terms of the angular coordinate in the orthogonal system,
namely~$\tht$; remember that~$\tp$ varies with~$\tht$ and~$s$ according
to~(\ref{eq:posi}).  Observe the scale factors are all positive
provided $0<r<1/\kappa$ and so the coordinate system is well defined
for unit radius pipes provided the non-dimensional centre line curvature
$\kappa<1$\,.  Let the velocity field, with components the axial
velocity~$u$, the radial velocity~$v$, and the angular velocity~$w$, be
denoted by
\begin{displaymath}
	\vel=u\vec e_s+v\vec e_r+w\vec e_\tht\,.
\end{displaymath}
Then, noting it is convenient to compute the viscous dissipation 
term via the vorticity (as does Tuttle~\cite[p548]{Tuttle90}),
\begin{displaymath}
	\nabla^2\vel=-\curl\vec\omega\,,\quad
	\vec\omega=\curl\vel\,,
\end{displaymath}
standard formulae apply for computing components of the Navier-Stokes 
equations~(\ref{eq:ns}) \cite[Appendix~B, e.g.]{Batchelor67}:
\begin{eqnarray}
	\omega_s & = & \frac{1}{r}\left( \D r{(rw)} -\D\tht v \right)\,,
	\label{eq:oms}  \\
	\omega_r & = & \frac{1}{rh_s}\left( \D\tht{(h_su)} -\D s{(rw)} 
	\right)\,,
	\label{eq:omr}  \\
	\omega_\tht & = & \frac{1}{h_s}\left( \D sv -\D r{(h_su)} \right)\,,
	\label{eq:omt}  \\
	0 & = & \frac{1}{h_s}\D sp +\frac{1}{r}\left( \D r{(r\omega_\tht)} 
	-\D\tht{\omega_r} \right)
	\nonumber\\&&{}
	+\re\left( \frac{u}{h_s}\D su +v\D ru +\frac{w}{r}\D\tht u 
	+\frac{uv}{h_s}\D r{h_s} +\frac{uw}{rh_s}\D\tht{h_s} \right)\,,
	\label{eq:nss}  \\
	0 & = & \D rp +\frac{1}{rh_s}\left( \D\tht{(h_s\omega_s)} -\D 
	s{(r\omega_\tht)} \right)
	\nonumber\\&&{}
	+\re\left( \frac{u}{h_s}\D sv +v\D rv +\frac{w}{r}\D\tht v 
	-\frac{w^2}{r} - \frac{u^2}{h_s}\D r{h_s} \right)\,,
	\label{eq:nsr}  \\
	0 & = & \frac{1}{r}\D \tht p +\frac{1}{h_s}\left( \D s{\omega_r} 
	-\D r{(h_s\omega_s)} \right)
	\nonumber\\&&{}
	+\re\left( \frac{u}{h_s}\D sw +v\D rw +\frac{w}{r}\D\tht w 
	-\frac{u^2}{rh_s}\D\tht{h_s} +\frac{vw}{r} \right)\,,
	\label{eq:nst}  \\
	0 & = & \frac{1}{rh_s}\left( \D s{(ru)} +\D r{(rh_sv)} 
	+\D\tht{(h_sw)} \right)\,.
	\label{eq:cty}
\end{eqnarray}
These are solved with a fixed fluid flux and with zero velocity on the
pipe walls: $u=v=w=0$ on~$r=1$\,.  The computer algebra program in
Appendix~\ref{Sca} solves these equations iteratively.

There are some subtleties in the geometry of the coordinate system.  As
discussed by Zabielski~\cite[\S2.2]{Zabielski98a}, observe that because
of the twist in a helical pipe the axial unit vector~$\vec u$ is
\emph{not everywhere} tangent to the lines of helical symmetry---the
$s$-coordinate curves are not curves of helical symmetry.  Thus be
careful in interpreting cross-flow velocities $v$~and~$w$ because in
one view they will involve a small component of the relatively large
velocity along the lines of helical symmetry.  In an alternative
presented by Tuttle~\cite{Tuttle90}, the twist in the coordinate system
caused by torsion generates an effect similar to that caused by a
coordinate system rotating in time.  However, here we consider flow in
a generally curving pipe with no large scale symmetry, so the only
definite longitudinal direction is the local unit vector~$\vec u$ and
we thus discuss~$v$ and~$w$ as cross-flow velocities, as does Gammack
\& Hydon~\cite{Gammack01}.  Similarly, in helical symmetry one cannot
find a cross-section \emph{plane} normal to the lines of helical
symmetry \cite[p300]{Zabielski98a} so an arbitrary decision is needed.
As is conventional for helical pipes and as simplest for generally
curving pipes, we conventionally take a cross-section to be normal to
the centreline of the pipe.  In these cross-sections the pipe is
circular.

\subsection{Slow Stokes flow}
\label{SSstokes}

Solving the fluid equations using the computer algebra program in 
Appendix~\ref{Sca} I deduce the Stokes flow field, $\re=0$, is 
\begin{eqnarray}
	u & = & \omrr \left[2 +\kappa\rat{3}{2}r\cos\tp 
	+\kappa^2\rat{5}{8}r^2\cos2\tp -\kappa^2\rat{11}{48}(1-3r^2)\right]
	\nonumber\\&&\quad
	{}+\Ord{\kappa^3,\delta^2,\re}\,,
	\label{eq:toru}  \\
	v & = & \rat{1}{3}\omrr ^2
	\left[\cos\tp\,\kappa' +\sin\tp\,\kappa\tau\right]
	\nonumber\\&&{}
	+\rat1{96} r\omrr ^2 \left[ 
	(38+43\cos2\tp)\kappa\kappa' +43\sin2\tp\,\kappa^2\tau \right]
	\nonumber\\&&{}
	+\Ord{\kappa^3,\delta^2,\re}\,,
	\label{eq:sv}  \\
	w & = & \rat{1}{6}\omrr \left(2-r^2\right)
	\left[-\sin\tp\,\kappa'+\cos\tp\,\kappa\tau\right]
	\nonumber\\&&{}
	+\rat1{96}r\omrr  \left[ 
	(43-29r^2)(\cos2\tp\,\kappa^2\tau -\sin2\tp\,\kappa\kappa') 
	+(6-2r^2)\kappa^2\tau \right]
	\nonumber\\&&{}
	+\Ord{\kappa^3,\delta^2,\re}\,,
	\label{eq:sw}  \\
	{\bar p}' & = & -8+\rat{1}{6}\kappa^2+\Ord{\kappa^3,\delta^2,\re}\,,
	\label{eq:torp} \\
	p & = & \bar p -\rat{1}{3}r\left(1-3r^2\right)
	\left[\cos\tp\,\kappa'+\sin\tp\,\kappa\tau\right]
	\nonumber\\&&{}
	-\rat1{24}\left(5+4r^2-21r^4\right)\kappa\kappa'
	-\rat1{24}r^2\left(9-26r^2\right) \left[\cos2\tp\,\kappa\kappa' 
	+\sin2\tp\,\kappa^2\tau\right]
	\nonumber\\&&{}
	+\Ord{\kappa^3,\delta^2,\re}\,,
	\label{eq:sp}
\end{eqnarray}
where~$\delta$ is used to denote the order of magnitude of derivatives 
of the quantities varying slowly along the pipe.
For example, $\kappa'$ and $\tau=-\phi'$ are thus~$\Ord{\delta}$.

\begin{figure}[tbp]
	\centering
	\includegraphics{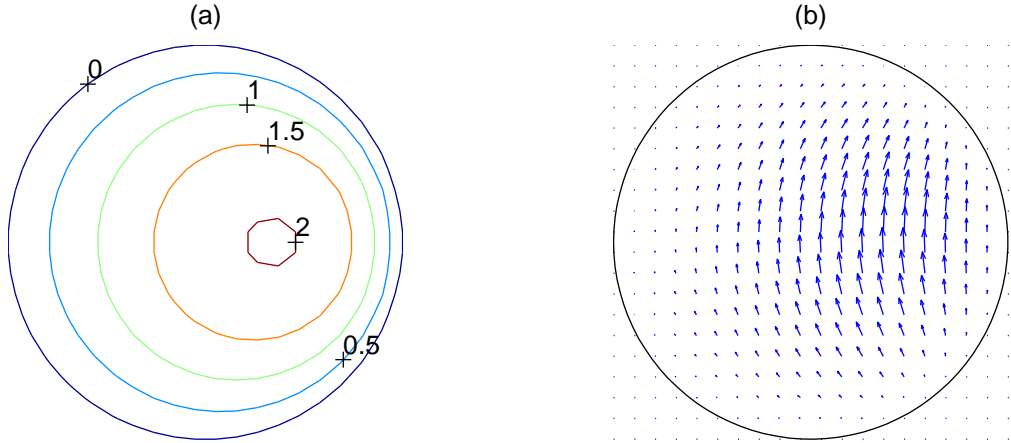} \caption{(a) contours of axial 
	velocity~$u$ of the viscously dominated Stokes flow in a helical 
	pipe with curvature~$\kappa=0.8$ (chosen so large to accentuate the 
	modifications); (b) corresponding torsion 
	induced cross-pipe fluid velocities to leading order in the 
	torsion.  The plots are evaluated from the asymptotic solution 
	with errors~$\Ord{\kappa^5}$.}
	\label{Fstokesq}
\end{figure}%
See that, for example, the Stokes flow in a torus 
($\kappa=\mbox{const}$ and $\tau=0$) is simply one of axial flow, see 
Figure~\ref{Fstokesq}(a), in an adjusted mean pressure gradient as all 
other components vanish.
The axial velocity maximum is shifted to the inside of the curve (to 
the right in Figure~\ref{Fstokesq}(a)) and is increased slightly.
In contrast to flows at significant Reynolds number, the pressure 
gradient around a curve is less than that in a straight pipe 
presumably because the bulk of the fluid travels a shorter path than 
the centre line---this agrees with Larrain~\cite{Larrain70} who used 
computer algebra to also find high order approximations to the flow in 
a coiled pipe.

The cross-pipe velocities in a helical pipe are indicated in
Figure~\ref{Fstokesq}(b) where the torsion induces velocities
proportional to those shown in the figure; the generally upwards
velocity matching the upwards twist of positive torsion.  Observe that
torsion, $\tau$, and variations in curvature, $\kappa'$, only affect
the cross-stream velocities and do not influence the axial velocity~$u$
to this order.  Conversely, observe that this Stokes flow does
\emph{not} have cross-pipe circulation---the strong viscosity
eliminates inertia.  Instead the curvature of the pipe just skews and
alters the velocity field.  For curvature $\kappa\neq 0$ the maximum of
the axial velocity~$u$ increases and moves towards the inner wall of
the pipe.  This last effect, though seemingly small even for the large
curvature of $\kappa=0.8$ used in Figure~\ref{Fstokesq}, is enough to
have a strong influence on the shear dispersion as seen
in~(\ref{eq:difad}).

\subsection{Laminar flow at finite Reynolds number}
\label{SSgen}

\begin{figure}[tbp]
	\centering
	\includegraphics{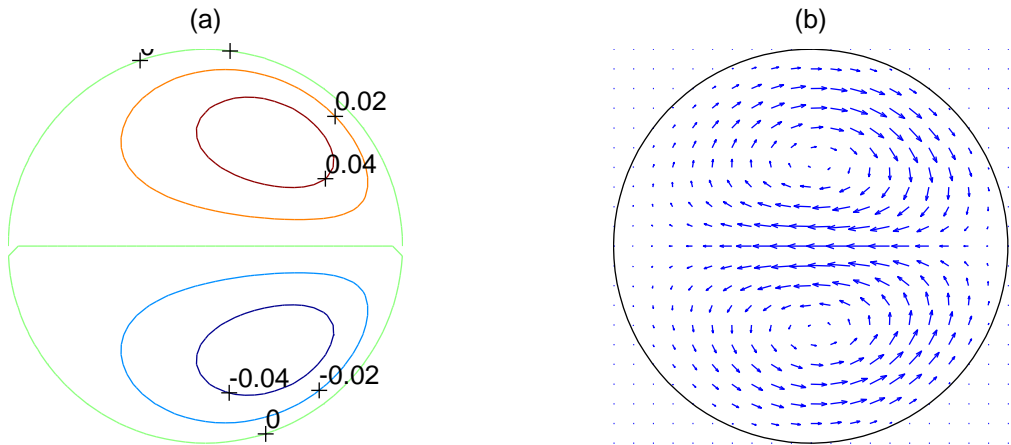} \caption{(a) axial velocity~$u$ contours
	of the inertial corrections to Poiseuille flow in a helical pipe
	with curvature $\kappa=0.5$ to leading order in the torsion~$\tau$
	of the helix; (b) corresponding inertia induced cross-pipe fluid
	velocities independent of the torsion.  The plots are evaluated
	from an asymptotic solution with errors~$\Ord{\kappa^5}$.}
	\label{Finertiaq}
\end{figure}%
Incorporating the terms representing the advection of fluid momentum
into the computer algebra program of Appendix~\ref{Sca} leads to
effects parametrised by the Reynolds number~$\re$.  I find that the
fluid fields given previously in~(\ref{eq:toru}--\ref{eq:sp}) are modified
by the addition of the following terms:
\begin{eqnarray}
	u & = & \cdots +\frac{\re}{144}r\omrr \left( 29+5r^2-3r^4 
	\right)\left[ \cos\tp\,\kappa'+\sin\tp\,\kappa\tau \right]
	\nonumber\\&&{}
	-\frac{\re^2}{1440}r\omrr\left( 19-21r^2+9r^4-r^6 
	\right)\cos\tp\,\kappa
	\nonumber\\&&{}
	+\frac{\re^3}{1814400}r\omrr \left( 2969 -4381r^2 +3249r^4 
	-1301r^6 
	\right.\nonumber\\&&\left.\quad{}
	+274r^8 -20r^{10} \right)
	\left[ \cos\tp\,\kappa' +\sin\tp\,\kappa\tau \right]
	+\Ord{\kappa^2,\delta^2}\,,
	\label{eq:nsu}  \\
	v & = & \cdots -\frac{\re}{72}\omrr ^2\left(4-r^2\right) 
	\cos\tp\,\kappa
	\nonumber\\&&{}
	+\frac{\re^2}{8640}\omrr^2\left( 13-15r^2+7r^4-r^6 \right)
	\left[ \cos\tp\,\kappa'+\sin\tp\,\kappa\tau \right]
	\nonumber\\&&{}
	+\Ord{\kappa^2,\delta^2}\,,
	\label{eq:nsv}  \\
	w & = & \cdots +\frac{\re}{72}\omrr \left( 4-23r^2+7r^4 
	\right)\sin\tp\,\kappa
	\nonumber\\&&{}
	+\frac{\re^2}{8640}\omrr\left( 13 -224r^2 +266r^4 -124r^6
	\right.\nonumber\\&&\left.\quad{}
	+17r^8 \right)
	\left[-\sin\tp\,\kappa'+\cos\tp\,\kappa\tau\right]
	+\Ord{\kappa^2,\delta^2}\,,
	\label{eq:nsw}  \\
	p & = & \cdots -\frac{\re}{3}r\left( 9-6r^2+2r^4 \right)\cos\tp\,\kappa
	\nonumber\\&&{}
	+\frac{\re^2}{2160}r\left( 101 -120r^2 +90r^4 -30r^6 +3r^8 \right)
	\left[ \cos\tp\,\kappa'+\sin\tp\,\kappa\tau \right]
	\nonumber\\&&{}
	+\Ord{\kappa^2,\delta^2}\,,
	\label{eq:nsp}
\end{eqnarray}
where ``$\cdots$'' denote the terms already given for Stokes flow
in~(\ref{eq:toru}--\ref{eq:sp}).  The modifications to the cross-pipe
velocities that are proportional to $\re\kappa(=\dn^2/4\re)$, plotted
in Figure~\ref{Finertiaq}, agree with those of the Dean flow as used by
Johnson~\cite[p330]{Johnson86}, in their work on dispersion.  The
cross-pipe velocity field exhibits circulation across the pipe induced
by the pipe curvature because of fluid inertia.  The term in the axial
velocity~$u$ proportional to $\kappa \re^2=\dn^2/4$ also agrees with
that of Johnson \& Kamm.  \marginpar{to this order, only modified by
the torsion: an upwards twisting causing the velocity peak to
apparently shift upwards.} These terms in the velocity fields are those
previously found for the ``loosely coiled limit'' \cite[p467]{Berger83}
when curvature~$\kappa$ is negligible by itself but the Dean number
$\dn=2\sqrt\kappa\re$ is significant.  The above expressions appear to
agree precisely with the expressions~(58--61) carefully obtained by
Tuttle~\cite{Tuttle90} for low Reynolds number flow in a helical
pipe---the only differences lie in various factors of two due to the
different non-dimensionalisation and because my angular velocity~$w$ is
in a ``space-centred'' coordinate system whereas Tuttle's~$\Phi$ is
``body centred''.\footnote{Although the components of the velocity
proportional to curvature~$\kappa$ and the leading order terms
in~$\re\kappa'$ reduce to those of
Murata~\cite[Eqn.~(21--22)]{Murata76}, in their case of a sinusoidal
centreline, the terms in~$\re^2$ are different in detail to those of
Murata~\cite[Eqn.~(22)]{Murata76}, as is the pressure.  Terms in
curvature~$\kappa$ in the above velocity field agree with those of
Pedley~\cite[Eqn.~(4.13)]{Pedley80}, and terms in the
gradient~$\kappa'$ with highest power of Reynolds number~$\re$ also
agree \cite[(4.18--19)]{Pedley80} except for the axial velocity~$u$.
The above expressions also agree with the small Dean number expansion
derived by Kao~\cite[p341]{Kao87}, for flow in a helix except for
his~$w_2$ which does not match my expression~(\ref{eq:nsu}) for~$u$.  }
Lastly, the components of the physical
fields~(\ref{eq:toru}--\ref{eq:nsp}) involving $\sin\theta$ and
$\cos\theta$ cross-sectional structures agree with those of Hammack \&
Hydon~\cite[p363]{Gammack01}, but their formulae have none of the
components in $\sin2\theta$, $\cos2\theta$, nor a modification to the
mean pressure gradient as in~(\ref{eq:meanpgrad}).  Observe in all of the
formulae~(\ref{eq:sv}--\ref{eq:nsp}) that torsion coupled to curvature,
$\kappa\tau$, appears to have the same effect as longitudinal gradients
of curvature, $\kappa'$, but the fluid fields are rotated in angle
by~$90^{\circ}$.  The above expressions are the first to combine
torsion and general variations in curvature.

With fixed fluid flux, the mean pressure gradient, to one higher order
in curvature than the fields above, is
\begin{eqnarray}
	{\bar p}'&=&-8
	+\left( \rat{1}{6} -\rat{11}{540}\re^2 
	-\rat{1541}{32659200}\re^4 \right)\kappa^2
	\nonumber\\&&{}\quad
	+\left( \rat{23}{80}\re +\rat{1433}{241920}\re^3 
	+\rat{6191}{410572800}\re^5 \right)\kappa\kappa'
	+\Ord{\kappa^3,\delta^2}\,.
	\label{eq:meanpgrad}
\end{eqnarray}
\begin{figure}[tbp]
	\centering
	\includegraphics{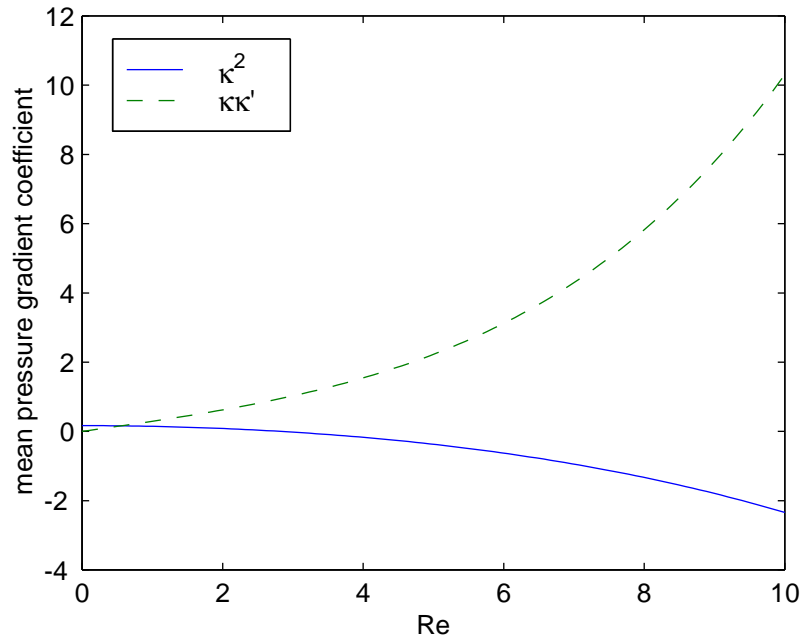}
	\caption{coefficients in the expression~(\ref{eq:meanpgrad}) for 
	the mean pressure gradient as a function of the centreline 
	curvature~$\kappa$ and its gradient~$\kappa'$.}
	\label{fig:meanpgrad}
\end{figure}%
See from the coefficients plotted in Figure~\ref{fig:meanpgrad}, that 
although the pressure gradient is lessened by curvature for low 
Reynolds number, for Reynolds number approximately~$\re>3$ there is an 
increased pressure gradient loss in a curving pipe.  This is 
attributed to the greater mixing caused by the induced cross-pipe 
circulation.  Also see that a region of tightening curvature, 
increasing~$\kappa^2$, has a lesser drop in pressure gradient relative 
to that for a toroidal pipe, whereas conversely a lessening in the 
curvature, decreasing~$\kappa^2$, has a higher drop in the pressure 
gradient.  This effect is interpreted as a ``memory'' in the 
mean pressure gradient of the upstream conditions: retaining just the 
highest order terms in~$\re$ the mean pressure gradient 
\begin{eqnarray}
	{\bar p}'&\approx& -8
	-\rat{1541}{32659200}\re^4 \kappa^2
	+\rat{6191}{821145600}\re^5 2\kappa\kappa'
	\nonumber\\
	&\approx& 
	-8-\rat{1541}{32659200\times16} 
	\left. \dn^4 \right|_{s-{4337}\re/{271216}}\,,
	\label{eq:meanpg}
\end{eqnarray}
where the evaluation of the Dean number at an effective distance
upstream of approximately~$\re/6$ pipe radii seems to show the typical
distance necessary for the fluid flow to develop in order to accord
with the curvature of the pipe.  This agrees qualitatively with
experiments on a pipe with a finite bend as discussed by Berger
\etal~\cite[p494]{Berger83}, where the influence of the bend on the
mean pressure gradient extends far downstream.  This upstream memory
also matches nicely with the commonly quoted distance,
$l_s\approx\frac14 a\re$ \cite{Fargie71} and \cite[p488]{Berger83},
required for flow entering a straight pipe to become fully developed,
and with the observation by Murata~\cite[\S4]{Murata76}, that the flow
in a sinusoidally bent pipe has a lag in its adaptation to the local
conditions, the lag increasing with increasing Reynolds number.

The above formulae for both viscous and inertia effects of curvature
and torsion are only of low-order.  Computer algebra computes the
velocity field to as high an order as is necessary for the demands of
modelling the dispersion in the pipe, described in the next section.
Van Dyke \cite[p475]{Berger83} has shown the asymptotic expansions of
the fluid flow field converge for Dean number $\dn<96.8/4\sqrt2=17.1$
(for negligible~$\kappa$ but finite~$\dn$).  However, I have not
explored this issue as here we are primarily concerned with the
advection-diffusion model~(\ref{eq:ad}) of the dispersion and \emph{its}
asymptotic approximations.

\section{Advection-dispersion along the curving pipe}
\label{Sdisp}

Having determined the fluid flow within the pipe, I now address the 
advection and longitudinal dispersion within the pipe.  We  
solve the advection-diffusion equation~(\ref{eq:ad}) for the evolving 
concentration $c(s,r,\tht,t)$ of contaminant within the fluid.  
Writing the concentration in terms of the cross-pipe average~$C(s,t)$ 
and its derivatives I use centre manifold techniques to construct  
the Taylor model~(\ref{eq:tay}), and its higher order generalisations, 
of the advection-dispersion in the bent and twisted pipe.

\subsection{The concentration field within the pipe}
\label{SScfield}

Using the coordinate system described in \S\ref{SScoord}, the 
advection-diffusion equation~(\ref{eq:ad}) for the evolution of the 
concentration of the contaminant is 
\begin{eqnarray}
	&&
	\D tc +\pe\left( \frac{u}{h_s}\D sc +v\D rc +\frac{w}{r}\D\tht c \right)
	\nonumber\\&&
	=\frac{1}{rh_s}\left[ \D s{}\left(\frac{r}{h_s}\D sc\right) 
	+\D r{}\left(rh_s\D rc\right) 
	+\D\tht{}\left(\frac{h_s}{r}\D\tht c\right) \right]\,.
	\label{eq:ceq}
\end{eqnarray}
This is solved with no flux through the circular walls of the pipe: 
$\partial c/\partial r=0$ on~$r=1$\,.

The computer algebra program (Appendix~\ref{Sca}) simultaneously 
determines from the contaminant conservation equation~(\ref{eq:ceq}) 
the dynamics on the low-dimensional centre manifold, namely the Taylor 
model~(\ref{eq:tay}).  Over a cross-pipe diffusion time the 
concentration field evolves to be, for example, approximately
\begin{eqnarray}
	c&=&C-\pe\D sC\frac{1}{24}\left(2-6r^2+3r^4\right)
	+\pe\D sC\kappa\cos\tp\,\left[
	-\frac{1}{6}\left(4r-3r^3+r^5\right)
	\right.\nonumber\\&&\quad\left.\quad{}
	+\frac{\re^2}{172800}\left(256r -285r^3 +200r^5 -75r^7 +15r^9 
	-r^{11}\right)
	\right.\nonumber\\&&\quad\left.\quad{}
	+\frac{\sigma\re^2}{34560}\left(68r -120r^3 +130r^5-75r^7 
	+21r^9 -2r^{11}\right)
	\right]
	\nonumber\\&&\quad{}
	+\Ord{\kappa^2,\delta^2}\,.
	\label{eq:sc}
\end{eqnarray}
To this order neither torsion nor gradients of curvature affect the
concentration field within the pipe, but this is not surprising as one
order of~$\delta$ is counted in~$\partial C/\partial s$ leaving no
scope for derivatives of the curvature~$\kappa$ to be involved in the
above terms.  Expressions for the concentration field to the next order
in either the curvature~$\kappa$ or longitudinal derivatives~$\delta$
are algebraically formidable and are not recorded.

From such expressions, the computer algebra determines the mean flux of 
contaminant through a cross-section of the pipe:
\begin{displaymath}
	F(C,s)=\overline{uc-\frac{1}{h_s}\D sc}\,,
\end{displaymath}
where the overbar denotes the average over a cross-section.  Then by 
conservation of contaminant, the model for the evolution of the 
contaminant is known to follow $C_t=-F_s$\,.  Using the flux~$F$ like 
this I determine the right-hand side of the model to one order higher 
in axial derivatives, $\delta$, than would otherwise be possible.

\afivepage

\subsection{Arbitrary curvature causes upstream memory}

The error of~$\Ord{\delta}$ in the shear dispersion coefficient given 
by~(\ref{eq:difad}) encompasses modifications due to both torsion~$\tau$ 
and to variations along the pipe in the curvature~$\kappa$.  The 
torsion only affects the dispersion coefficient at 
$\Ord{\kappa^2\tau^2}$ and so does not show up in~(\ref{eq:difad}).

Variations in curvature along the pipe ($\kappa'\neq 0$) cause the 
effective dispersion coefficient to become, using~$\rr$ to denote the 
scaled Reynolds number~$\re/10$ and remembering $\pe=\re\sigma$,
\begin{subequations}
\begin{eqnarray}
	D &\!\!\!=\!\!\! & 
    1 +\frac{1}{4}\kappa^2 +\frac{7}{48}\left(\kappa\kappa'\right)' 
	 -\frac{7}{96}\left(\kappa^2\tau^2+{\kappa'}^2\right) 
	\label{eq:deffa}\\&&{}
	+\frac{\pe^2}{48}\left[ 1 +\kappa^2\left( 
	-\rat{64225}{171072}\rr^4\sigma^2 +\rat{2995}{24192}\rr^4
	-\rat{36335}{12096}\rr^2  +\rat{863}{120}\right) \right]
	\label{eq:deffb}\\&&{} 
	+\pe\kappa\kappa'\left[
	\rr^{6} \left(\rat{9050586625}{26900729856} \sigma^{4}
	+\rat{246093875}{1630347264} \sigma^{3}
	-\rat{6234774125}{53801459712} \sigma^{2}
	-\rat{1760495125}{40351094784} \sigma\right)
	\right.\nonumber\\&&\quad\left.{}
	+\rr^{4} \left(-\rat{1068925}{2322432} \sigma^{3}
	+\rat{33738035}{20901888} \sigma^{2}
	+\rat{2310385}{4478976} \sigma\right)
	\right.\nonumber\\&&\quad\left.{}
	+\rr^{2} \left(-\rat{383695}{96768} \sigma^{2}
	+\rat{19465}{12096} \sigma
	+\rat{985}{12096}\right)
	\right.\nonumber\\&&\quad\left.{}
	-\rat{13}{32}\right]
	\label{eq:deffc}\\&&
	+\pe^2(\kappa^2\tau^2+{\kappa'}^2)\left[
	\rr^{6} \left(\rat{2542365125}{58692501504} \sigma^{4}
	-\rat{1039029345155}{126541033242624} \sigma^{2}
	-\rat{5542735225}{3515028701184}\right)
	\right.\nonumber\\&&\quad\left.{}
    +\rr^{4} \left(\rat{14791164485}{33108590592} \sigma^{2}
	+\rat{739414405}{11036196864}\right)
	\right.\nonumber\\&&\quad\left.{}
    +\rr^{2} \left(-\rat{7181}{9072} \sigma^{2}
	+\rat{7619671}{41803776}\right)
	\right.\nonumber\\&&\quad\left.{}
    -\rat{5357}{55296}
	\right]
	\label{eq:deffd}\\&&
% 	+\pe\rr{\kappa'}^2\left[
% 	\rr^{6} \left(-\rat{2744107466875}{9038645231616} \sigma^{5}
% 	-\rat{3007270625375}{9038645231616} \sigma^{4}
% 	+\rat{412584032375}{10545086103552} \sigma^{3}
% 	\right.\right.\nonumber\\&&\qquad\left.\left.{}
% 	+\rat{2879131496575}{23726443732992} \sigma^{2}
% 	+\rat{2629472875}{282457663488} \sigma\right)
% 	\right.\nonumber\\&&\quad\left.{}
% 	+\rr^{4} \left(\rat{318184675}{306561024} \sigma^{4}
% 	+\rat{211750175}{788299776} \sigma^{3}
% 	-\rat{9765145925}{16554295296} \sigma^{2}
% 	+\rat{516914675}{2759049216} \sigma\right)
% 	\right.\nonumber\\&&\quad\left.{}
% 	+\rr^{2} \left(\rat{415705}{516096} \sigma^{3}
% 	-\rat{3081595}{1548288} \sigma^{2}
% 	+\rat{1048765}{2985984} \sigma
% 	-\rat{16835}{373248}\right)
% 	\right.\nonumber\\&&\quad\left.{}
% 	+\rat{469}{2304} \sigma-\rat{191}{1728}
% 	\right]
% 	\label{eq:deffe}\\&&
	+\pe\rr(\kappa\kappa')'\left[
	\rr^{6} \left(-\rat{2219783253125}{3012881743872} \sigma^{5}
	-\rat{3007270625375}{9038645231616} \sigma^{4}
	+\rat{7670650920025}{63270516621312} \sigma^{3}
	\right.\right.\nonumber\\&&\qquad\left.\left.{}
	+\rat{2879131496575}{23726443732992} \sigma^{2}
	+\rat{396673566125}{15817629155328} \sigma\right)
	\right.\nonumber\\&&\quad\left.{}
	+\rr^{4} \left(\rat{318184675}{306561024} \sigma^{4}
	-\rat{3159503125}{752467968} \sigma^{3}
	-\rat{9765145925}{16554295296} \sigma^{2}
	-\rat{2663242675}{5518098432} \sigma\right)
	\right.\nonumber\\&&\quad\left.{}
	+\rr^{2} \left(\rat{40508065}{4644864} \sigma^{3}
	-\rat{3081595}{1548288} \sigma^{2}
	-\rat{3844625}{2612736} \sigma
	-\rat{16835}{373248}\right)
	\right.\nonumber\\&&\quad\left.{}
	+\rat{32413}{27648} \sigma-\rat{191}{1728}
	\right]
	\label{eq:defff}
	+\Ord{\kappa^4,\delta^3}\,,
\end{eqnarray}
	\label{eq:deff}
\end{subequations}
In this large but comprehensive expression observe:
\begin{description}
	\item[(\ref{eq:deffa})] gives the molecular diffusivity along the 
	pipe in the presence of the bending and twisting of the pipe when 
	there is no flow;

	\item[(\ref{eq:deffb})] gives the usual shear enhanced dispersion in
	a straight pipe, $\pe^2/48$, modified by the leading order
	(quadratic) effects of pipe curvature (these were the terms of the
	shear dispersion discussed in the Introduction~(\ref{eq:difad});

	\item[(\ref{eq:deffc})] gives the leading order effects on the 
	dispersion due to variations in curvature along the pipe;

	\item[(\ref{eq:deffd})] gives the leading order effect of torsion on 
	the dispersion, namely quadratic but moderated by the 
	multiplication by~$\kappa^2$;

	\item[(\ref{eq:deffd}--\ref{eq:defff})] through
	${\kappa'}^2$~and~$\kappa''$ terms, gives the second order effects
	of the variations in curvature.
	
	It is intriguing to see that the effects of torsion and second 
	order gradients of curvature factorise as shown 
	in~(\ref{eq:deffd}--\ref{eq:defff}).  I suggest the reason 
	for this factorisation is due to two effects: firstly, upstream 
	``memory'' of the dispersion, to be discussed later, involves
		\begin{displaymath}
		\left.\kappa^2\right|_{s-\xi}=\kappa^2 
		-2\xi\kappa\kappa'+\xi^2(\kappa\kappa')'+\Ord{\xi^3}
	\end{displaymath}
	which may explain the appearance of the combination 
	$(\kappa\kappa')'$; and secondly, curvature gradients and torsion, 
	$\kappa'$ and~$\kappa\tau$ respectively, both create the same but 
	orthogonal structures in the fluid flow as commented 
	after~(\ref{eq:nsu}--\ref{eq:nsp}).

\end{description}

\begin{figure}[tbp]
	\centering
	\includegraphics{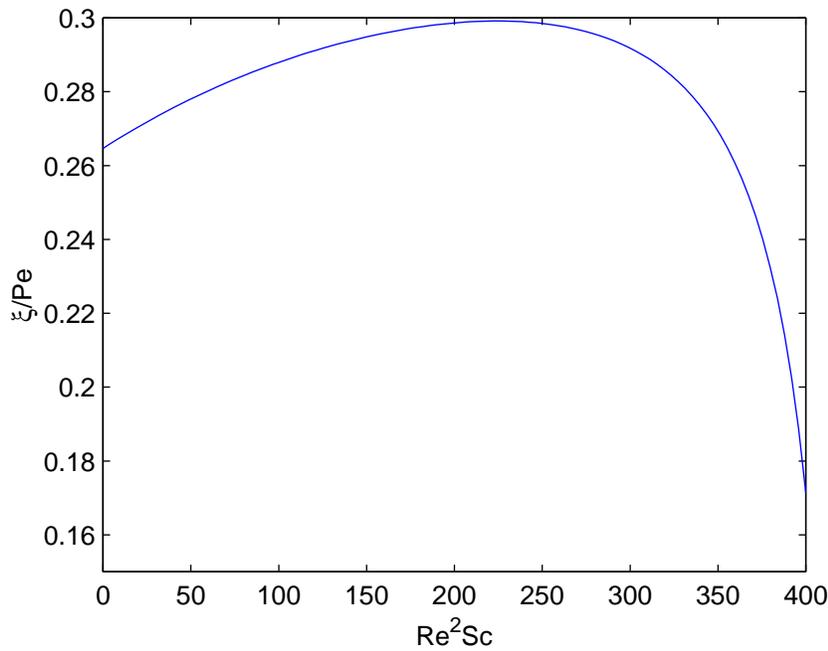} \caption{in a generally curving 
	pipe the effective dispersion has the value appropriate to the 
	curvature a distance~$\xi$ upstream.}
	\label{Fupstream}
\end{figure}

\paragraph{For large Schmidt number~$\sigma$} (typical for material
dispersion in liquids) there are two distinguished limits of the above
expression for the effective dispersion coefficient, the second being a
subset of the first.
\begin{itemize}
	\item Firstly, for large Schmidt number~$\sigma$ the highest powers
	of~$\sigma$ dominate.  However, in various subexpressions they
	appear in combination with the Reynolds number~$\re=10\rr$.  Thus
	there is a distinguished limit with large~$\sigma$ and small~$\re$
	in which $\re^2\sigma$ is of order~$1$\,.  In terms of the
	magnitude~$\delta$ of the slow axial variations, an appropriate
	scaling is that the fluid flow is slow, $\re\sim\delta$, the
	Schmidt number large enough, $\sigma\sim1/\delta^2$, so that the
	Peclet number is also large, $\pe\sim1/\delta$, then the effective
	diffusion coefficient is large, $D\sim 1/\delta^2$.  Using these
	orders of magnitude, introducing the order~$1$ parameter
	$\repe=\re\pe/100=\re^2\sigma/100$ and evaluating fractions, the
	leading order terms in the dispersion coefficient~(\ref{eq:deff})
	are
	\begin{eqnarray}
		D & \!\!\!\approx\!\!\! & .02083\,\pe^2
		+(.1498-.007821\,\repe^2)\pe^2\kappa^2
		\nonumber\\&&{}
		+(-.03965-.004603\,\repe +.003364\,\repe^2)\pe^3\kappa\kappa'
		\nonumber\\&&{}
		+(- .007916 + .0004332\,\repe^2)\pe^4(\kappa^2\tau^2+{\kappa'}^2)
		\nonumber\\&&{}
% 		+(.0008055 + .001038\,\repe - .0003036\,\repe^2)\pe^4{\kappa'}^2
% 		\nonumber\\&&{}
		+(.008721 + .001038\,\repe - .0007368\,\repe^2)\pe^4(\kappa\kappa')'
		\,.
		\label{eq:defFa}
	\end{eqnarray}

	\item  Secondly, for a typical Schmidt number~$\sigma$ bigger than 
	$10^3$ or so, and for any flow with Reynolds 
	number~$\re$ bigger than about~$1$, then the parameter~$\repe$ will 
	be bigger than about~$10$ and the above expression~(\ref{eq:defFa}) 
	will be dominated by the quadratic powers in~$\repe$.  That is, the 
	dispersion coefficient
		\begin{eqnarray}
		{D}&\!\!\!\approx\!\!\!& \left(\frac{\pe}{10}\right)^2
		\left\{2.083 
		+\left(\frac{\re^2\sigma}{100}\right)^2\left[ -0.7821\,\kappa^2 
		+3.364\,\frac\pe{10}\kappa\kappa' 
		\vphantom{\left(\frac{\pe}{10}\right)^2}
		\right.\right.\nonumber\\&&\left.\left.{}
		+\left(\frac{\pe}{10}\right)^2\left( 4.332\,(\kappa^2\tau^2+{\kappa'}^2)
% 		- 3.036\,{\kappa'}^2 
		- 7.368\,(\kappa\kappa')' \right)  \right] \right\}.
		\label{eq:defFb}
	\end{eqnarray}
\end{itemize}
We noted in~(\ref{eq:meanpg}) that the mean pressure gradient in the
fluid flow at any location was appropriate to the curvature some
distance upstream.  Similar memory effects are seen in the dispersion
coefficient.  The subexpression $-0.7821\,\kappa^2
+0.3364\,\pe\kappa\kappa'$ appearing in the first line
of~(\ref{eq:defFb}) is equivalent to simply $-0.7821\,\kappa^2$
evaluated at a distance $\xi=0.2151\,\pe$ upstream from any particular
location.  I do not attempt to complicate this memory effect any
further by trying to include the second order term $(\kappa\kappa')'$,
as there are a plethora of possibilities, but for the purposes of
discussion I assume both~$\kappa\kappa'$ and~$(\kappa\kappa')'$ terms
are attributable to upstream memory.  The ratio of the coefficients
of~$\kappa\kappa'$ and~$\kappa^2$ in~(\ref{eq:defFa}) similarly
quantify the upstream memory for low Reynolds number flows as shown in
Figure~\ref{Fupstream}.  Such memory effects in shear dispersion in
varying channels were first recognised by Smith~\cite{Smith83b}.

From the coefficient approximations~(\ref{eq:defFa})
and~(\ref{eq:defFb}) see that torsion and curvature gradients generally
enhance dispersion along the pipe except for low Reynolds numbers,
$\re^2\sigma<427.4$\,, when they make the dispersion coefficient
smaller.  However, the effect torsion has upon the dispersion
coefficient seems small because not only is the effect quadratic in the
torsion~$\tau$, it is also ameliorated by the multiplication by the
curvature squared.  However, ignoring the~$\kappa\kappa'$ terms and
noting that $\kappa'=\kappa(\log\kappa)'$, I write~(\ref{eq:defFb}) as
\begin{eqnarray}
		{D}&\!\!\!\approx\!\!\!& \left(\frac{\pe}{10}\right)^2
		\left\{2.083 
        \vphantom{\frac{\pe}{10}}\right.\nonumber\\&&\left.{}
		-0.7821\left(\frac{\kappa\re^2\sigma}{100}\right)^2\left[ 1 
		-\left(\frac{\pe}{10}\right)^2 5.539\,(\tau^2+{(\log\kappa)'}^2)
	    \right] \right\}.
	\label{eq:defftor}
\end{eqnarray}
% \begin{displaymath}
% 		{D}\approx \left(\frac{\pe}{10}\right)^2
% 		\left\{2.083 
% 		-0.7821\left(\frac{\kappa\re^2\sigma}{100}\right)^2\left[ 1 
% 		-\left(\frac{\pe}{10}\right)^2 5.539\,(\tau^2+{(\log\kappa)'}^2)
% 	    \right] \right\}.
% \end{displaymath}
This suggests that torsion, or proportional gradients of curvature,
greater than about~$4/\pe$ may cause the dispersion coefficient~$D$ to
increase with curvature~$\kappa$, instead of decreasing.  That torsion
could eliminate the increased mixing due to secondary circulations
seems unlikely so I predict higher order terms in the torsion~$\tau$
would limit its influence on the dispersion.

\begin{figure}[tbp]
	\centering
	\includegraphics[width=0.8\textwidth]{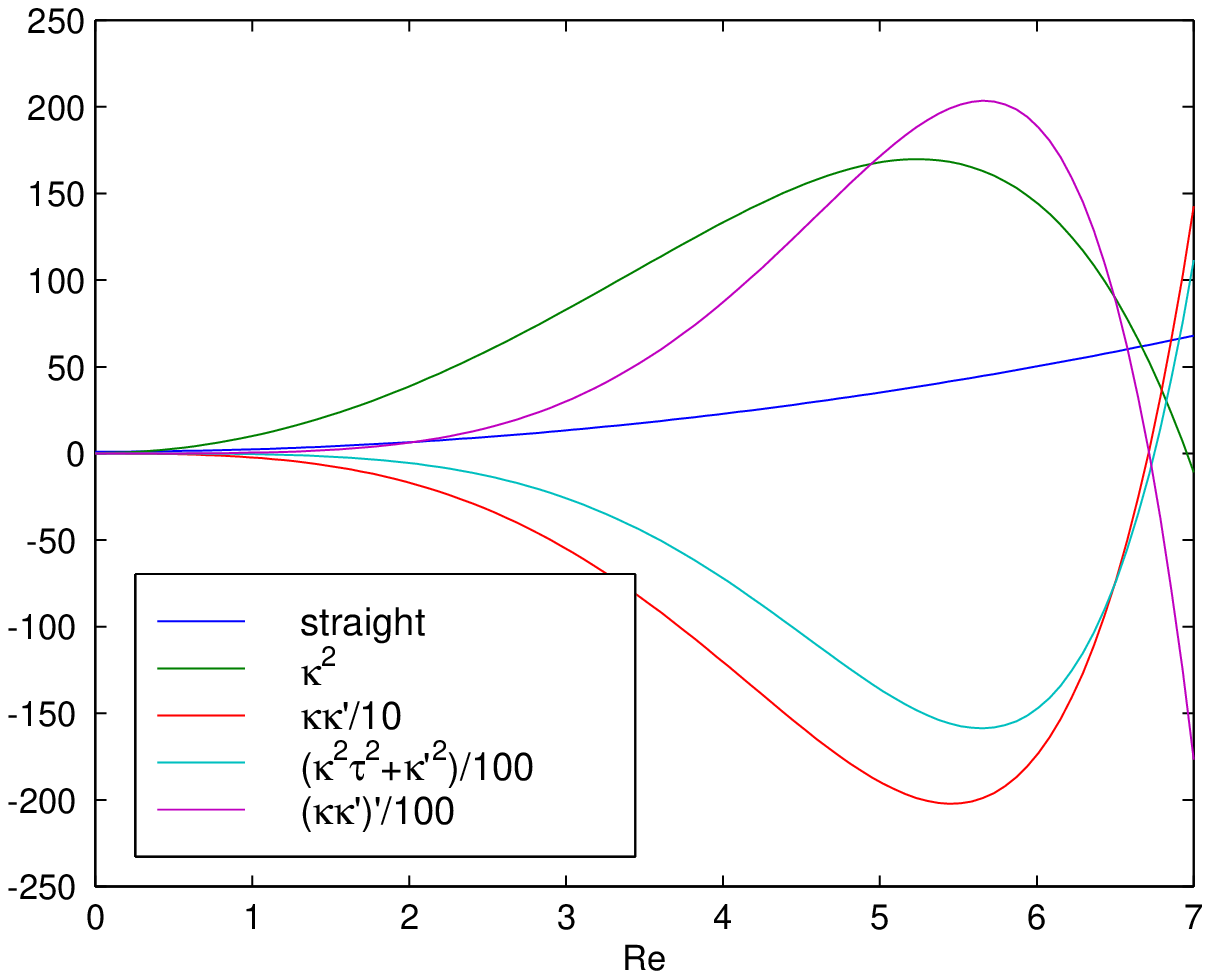}
	\caption{coefficients of the dispersion coefficient~$D$ for water at 
	$15^\circ\mbox{C}$ given by~(\ref{eq:deffw}).  For higher Reynolds 
	number the highest powers in~$\re$ dominate the coefficients.}
	\label{fig:dwater}
\end{figure}
\paragraph{For the dispersion of heat in water} with a Prandtl number of 
$\sigma=8.1$ at~$15^\circ\mbox{C}$ the dispersion 
coefficient~(\ref{eq:deff}) reduces to
\begin{eqnarray}
	\dauq&&D  =  1.3669\,\re^{2}+1
		\nonumber\\\dauq&&{}
	+\kappa^{2} \left(-.003350\,\re^{6}-.04106\,\re^{4}
	    +9.830\,\re^{2}+.25\right)
		\nonumber\\\dauq&&{}
	+\kappa \kappa' \left(.01232\,\re^{7}-.1090\,\re^{5}
	    -2.01\,\re^{3}-3.291\,\re\right)
		\nonumber\\\dauq&&{}
	+(\kappa^{2} \tau^{2} +{\kappa'}^{2}) \left(.01220\,\re^{8}+.1928\,\re^{6}
	    -33.95\,\re^{4}-6.356\,\re^{2}-.07292\right)
		\nonumber\\\dauq&&{}
% 	+{\kappa'}^{2} \left(-.009711\,\re^{8}+.3704\,\re^{6}
% 	    +2.432\,\re^{4}+1.246\,\re^{2}+.07292\right)
% 		\nonumber\\\dauq&&{}
	+(\kappa \kappa')'\left(-.02191\,\re^{8}+.1777\,\re^{6}
	    +36.39\,\re^{4}+7.602\,\re^{2}+.1458\right).
	\label{eq:deffw}
\end{eqnarray}
For Reynolds number~$\re>7$ the highest powers in~$\re$ dominate; see 
Figure~\ref{fig:dwater} for the dependence on smaller~$\re$.  Observe: 
for Reynolds number $\re>6.95$ curvature enhances the dispersion of 
heat and vice-versa; whereas for $\re>6.74$ torsion and curvature 
gradients reduce the dispersion and vice-versa.

\begin{figure}[tbp]
	\centering
	\includegraphics[width=0.8\textwidth]{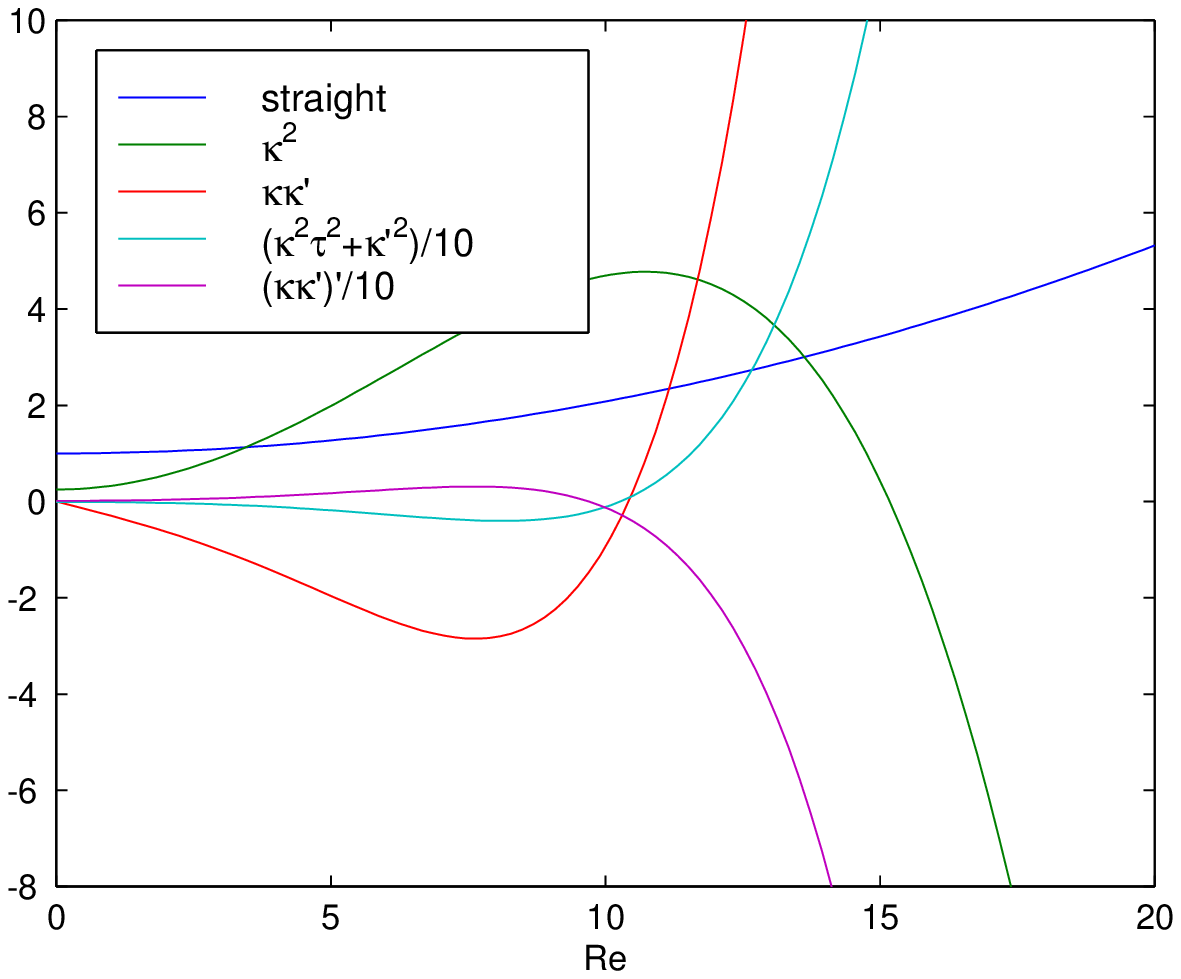}
	\caption{coefficients of the dispersion coefficient~$D$ for water at 
	$15^\circ\mbox{C}$ given by~(\ref{eq:deffair}).  For higher Reynolds 
	number the highest powers in~$\re$ dominate the coefficients.}
	\label{fig:dair}
\end{figure}
\paragraph{For the dispersion of heat in air} with a Prandtl number of 
$\sigma=0.71$ at~$15^\circ\mbox{C}$ and recalling that $\rr=\re/10$, 
the dispersion coefficient~(\ref{eq:deff}) reduces to
\begin{eqnarray}
	\dauq&&D  =  1.080\,\rr^{2} +1
		\nonumber\\\dauq&&{}
	+\kappa^{2} \left(-.07649\,\rr^{6}-3.244\,\rr^{4}
	    +7.767\,\rr^{2}+.25\right)
		\nonumber\\\dauq&&{}
	+\kappa \kappa' \left(.3979\,\rr^{7}+ 7.462\,\rr^{5}
	    - 5.871\,\rr^{3}- 2.925\,\rr\right)
		\nonumber\\\dauq&&{}
	+(\kappa^{2} \tau^{2}+{\kappa'}^{2}) \left(.3011\,\rr^{8}+ 15.48\,\rr^{6}
	    - 11.82\,\rr^{4}- 5.022\,\rr^{2}-.07292\right)
		\nonumber\\\dauq&&{}
% 	+{\kappa'}^{2} \left(- .4604\,\rr^{8}+ 1.500\,\rr^{6}
% 	    - 3.768\,\rr^{4}+ .2594\,\rr^{2}+.07292\right)
% 		\nonumber\\\dauq&&{}
	+(\kappa \kappa')'\left(- .7615\,\rr^{8}- 13.98\,\rr^{6}
	    + 8.055\,\rr^{4}+ 5.282\,\rr^{2}+.1458\right)\,.
	\label{eq:deffair}
\end{eqnarray}
For Reynolds number~$\re>15$ the highest powers in~$\re$ dominate; see 
Figure~\ref{fig:dair} for the dependence on smaller~$\re$.
Observe: for Reynolds number $\re>15.18$ curvature enhances the 
dispersion of heat and vice-versa; whereas for $\re>10.35$ torsion and 
curvature gradients reduce the dispersion and vice-versa.

\afivepage

\subsection{Skewness is very sensitive to curvature}
\label{sec:skew}

Computer algebra straightforwardly determines high order terms in the
advec\-tion-diffusion equation~(\ref{eq:ad}).  Chatwin~\cite{Chatwin70}
investigated the relatively slow approach to normality in shear
dispersion.  Here expect the variations in pipe curvature and torsion
to distort any normal profile.  Hence expect such variations to have a
large effect on skewness.

The third order modification to the Taylor model of dispersion is
\begin{equation}
	\D tC\approx -\pe\D sC+\D s{\ }\left(D\D sC\right)
	+\D s{\ }\left(E\DD sC\right)\,,
	\label{eq:third}
\end{equation}
where the skewness coefficient
\begin{subequations}
\begin{eqnarray}
	E&\!\!\!=\!\!\!&-\frac{\pe^3}{2880}
	\label{eq:skewa}\\&&{}
	+\pe\kappa^2\left[
\rr^{6} \left(\rat{3241338875}{107602919424} \sigma^{4}
+\rat{1104359125}{35867639808} \sigma^{2}\right)
-\rat{4085615}{20901888} \rr^{4} \sigma^{2}
	\right.\nonumber\\&&\left.\quad{}
+\rr^{2} \left(\rat{68855}{96768} \sigma^{2}
+\rat{985}{12096}\right)
-\rat{13}{32}
    \right]
	\label{eq:skewb}\\&&{}
	+\kappa\kappa'\left[
\rr^{8} \left(-\rat{5943982203125}{9038645231616} \sigma^{6}
-\rat{1183708683125}{2259661307904} \sigma^{5}
-\rat{12012557065375}{9038645231616} \sigma^{4}
	\right.\right.\nonumber\\&&\left.\left.\qquad{}
-\rat{4304114889625}{5931610933248} \sigma^{3}
\right)
	\right.\nonumber\\&&\left.\quad{}
+\rr^{6} \left(\rat{1250362375}{1379524608} \sigma^{5}
+\rat{820657375}{64665216} \sigma^{4}
+\rat{14670875}{4138573824} \sigma^{3}
+\rat{1214048125}{4138573824} \sigma^{2}\right)
	\right.\nonumber\\&&\left.\quad{}
+\rr^{4} \left(-\rat{11414525}{290304} \sigma^{4}
+\rat{1917625}{193536} \sigma^{3}
-\rat{25984475}{5225472} \sigma^{2}
-\rat{84175}{93312} \sigma\right)
	\right.\nonumber\\&&\left.\quad{}
+\rr^{2} \left(\rat{17795}{1152} \sigma^{2}
-\rat{955}{432} \sigma\right)
+\rat{7}{12}
	\right]
	\label{eq:skewc}\\&&{}
	+\pe(\kappa^2\tau^2+{\kappa'}^2)\left[
	\rr^{8} \left(\rat{254564364353125}{1434131710083072} \sigma^{6}
-\rat{9130311425570375}{34419161041993728} \sigma^{4}
	\right.\right.\nonumber\\&&\left.\left.\qquad{}
-\rat{545957180588375}{29041167129182208} \sigma^{2}\right)
	\right.\nonumber\\&&\left.\quad{}
+\rr^{6} \left(\rat{9773515705025}{2259661307904} \sigma^{4}
+\rat{241252377733475}{759246199455744} \sigma^{2}\right)
	\right.\nonumber\\&&\left.\quad{}
+\rr^{4} \left(-\rat{1227346555}{114960384} \sigma^{4}
-\rat{24438946685}{16554295296} \sigma^{2}
-\rat{159535}{2128896}\right)
	\right.\nonumber\\&&\left.\quad{}
+\rr^{2} \left(\rat{1325215}{258048} \sigma^{2}
-\rat{118243}{580608}
\right)
+\rat{801}{2560}
	\right]
	\label{eq:skewd}\\&&{}
	+\pe(\kappa\kappa')'\left[
	\rr^{8} \left(-\rat{26098554271144375}{206514966251962368} \sigma^{6}
+\rat{10015284615625}{204875958583296} \sigma^{5}
	\right.\right.\nonumber\\&&\left.\left.\qquad{}
+\rat{83884205830882375}{206514966251962368} \sigma^{4}
+\rat{68444215116292625}{464658674066915328} \sigma^{3}
+\rat{44871973001125}{691456360218624} \sigma^{2}\right)
	\right.\nonumber\\&&\left.\quad{}
+\rr^{6} \left(\rat{71359651375}{6025763487744} \sigma^{5}
-\rat{516258977875}{111588212736} \sigma^{4}
-\rat{596762793925}{2875932573696} \sigma^{3}
	\right.\right.\nonumber\\&&\left.\left.\qquad{}
-\rat{2958969842375}{23007460589568} \sigma^{2}
-\rat{62332574575}{1977203644416} \sigma\right)
	\right.\nonumber\\&&\left.\quad{}
+\rr^{4} \left(\rat{48287105}{3784704} \sigma^{4}
-\rat{480287515}{172440576} \sigma^{3}
+\rat{4233379915}{2759049216} \sigma^{2}
+\rat{89672375}{459841536} \sigma
+\rat{159535}{2128896}\right)
	\right.\nonumber\\&&\left.\quad{}
+\rr^{2} \left(-\rat{23023685}{4644864} \sigma^{2}
+\rat{2964127}{4644864} \sigma
+\rat{501107}{2322432}\right)
-\rat{2257}{5760}
	\right]
	\label{eq:skewe}
	+\Ord{\kappa^4,\delta^3}\,.
\end{eqnarray}
	\label{eq:skew}
\end{subequations}
The skewness coefficient for flow in a straight pipe, $-\pe^3/2880$
from~(\ref{eq:skewa}), is well known \cite{Chatwin85}.  One outstanding
puzzle in the field of dispersion is that in rivers one observes
contaminant concentrations with long tails upstream (not downstream) 
\cite[e.g.]{Beer83}.
But theoretical models predict either only a weak enhancement of
upstream tails or more confoundedly, as the above negative skewness
coefficient implies for straight pipe flow, a weak downstream tail.
However, as we now see, curvature effects, presumably induced by the
secondary flows, significantly change the skewness coefficient thereby
enhancing the upstream tail of a contaminant release.  With the caveat
that this derivation is for pipes, not rivers, this is a qualitative
improvement in the theoretical model compared with observations.

% Thus the skewness coefficient $\pe^2/2880$ is sensitive to curvature 
% effects:
% \begin{itemize}
% 	\item  in slow flow it is reduced by a factor roughly 
% 	$1-20\kappa^2$;
% 
% 	\item  in liquids it is reduced by about 
% 	$1-0.87(\kappa\re^2\sigma/100)^2$;
% 
% 	\item  whereas in gases the $\pe\kappa^2$ term is a significant 
% 	modification.
% \end{itemize}

\paragraph{For large Schmidt number~$\sigma$} (typical for the
dispersion of material in liquids), and similar to the dispersion
coefficient~$D$, there are two distinguished limits of the above
expression for the skewness coefficient.
\begin{itemize}
	\item Firstly, in terms of the magnitude~$\delta$ of the slow 
	axial variations, the appropriate scaling is that the fluid flow 
	is slow, $\re\sim\delta$, the Schmidt number large enough, 
	$\sigma\sim1/\delta^2$, so that the Peclet number is also large, 
	$\pe\sim1/\delta$, then the skewness coefficient is large, $E\sim 
	1/\delta^3$.  Recalling the parameter $\repe =\re\pe/100 
	=\re^2\sigma/100$ and evaluating fractions, the leading order 
	terms in the skewness coefficient~(\ref{eq:skew}) are
\begin{eqnarray}
	E & \approx & -\frac{\pe^3}{2880}\left[1 
	-\kappa^2\left(.8675 \,\repe^{2} +20.49\right)
	\right.\nonumber \\&  & \left.\quad{}
	 +\pe\kappa\kappa'\left(.1894 \,\repe^{2}-.2610 \,\repe+11.32\right)
	\right.\nonumber \\&  & \left.\quad{}
	 +\pe^2(\kappa^2 \tau^2+{\kappa'}^2) \left(-.05112\, \repe^{2}+3.075\right)
	\right.\nonumber \\&  & \left.\quad{}
	 +\pe^2(\kappa\kappa')'\left(.03640\, \repe^{2}-.003411\, \repe-3.674\right)
     \right]
	\label{eq:eeff}
\end{eqnarray}

	\item Secondly, for a typical Schmidt number~$\sigma$ bigger than
	$10^3$ or so, and for any flow with Reynolds number~$\re$ bigger
	than about~$2$, then the parameter~$\repe$ will be bigger than
	about~$40$ and the above skewness coefficient~(\ref{eq:eeff}) is
	dominated by the quadratic powers in~$\repe$:
		\begin{eqnarray}
		{E}&\approx& \frac{\pe^3}{2880}
		\left\{-1 
		+\left(\frac{\re^2\sigma}{100}\right)^2\left[0.8675\,\kappa^2 
		-1.894\,\frac\pe{10}\kappa\kappa' 
		\right.\right.\nonumber\\&&\left.\left.{}
		+\left(\frac{\pe}{10}\right)^2\left( 5.112\,(\kappa^2\tau^2+{\kappa'}^2)
% 		- 3.036\,{\kappa'}^2 
		- 3.640\,(\kappa\kappa')' \right)  \right] \right\}.
        \label{eq:eeffb}
	\end{eqnarray}
	In this regime, even small curvature, through the~$\kappa^2$ term,
	will cause the skewness coefficient to become positive, possibly
	large, and so lead to concentration tails upstream (qualitatively
	as observed in rivers).  Torsion in the pipe leads to the same
    upstream tails.
\end{itemize}

Recognise another upstream memory effect.  The subexpression
$0.8675\,\kappa^2 -0.1894\,\pe\kappa\kappa'$ appearing in the first
line of~(\ref{eq:eeffb}) is equivalent to simply $0.8675\,\kappa^2$
evaluated at a distance $\xi=0.1092\,\pe$ upstream from any particular
location.  This upstream memory is approximately half that of the
dispersion coefficient.

\paragraph{For the dispersion of heat in water} with a Prandtl number of 
$\sigma=8.1$ at~$15^\circ\mbox{C}$ the skewness 
coefficient~(\ref{eq:skew}) reduces to 
\begin{eqnarray}
	&&E  =  -0.1845\,\re^3
		\nonumber\\&&{}
+\kappa^2\left(
-3.291\,\re
+3.788\,\re^3
-.01039\,\re^5
+.001067\,\re^7\right)
		\nonumber\\&&{}
+\kappa\kappa'\left(.5833
+9.956\,\re^2
-16.43\,\re^4
+.08625\,\re^6
-.002101\,\re^8\right)
		\nonumber\\&&{}
+(\kappa^2\tau^2+{\kappa'}^2)\left(
2.534\,\re
+27.28\,\re^3
-37.30\,\re^5
+.1509\,\re^7
+.003968\,\re^9\right)
		\nonumber\\&&{}
+(\kappa\kappa')' \left(
-3.174\,\re
-25.91\,\re^3
+43.37\,\re^5
-.1589\,\re^7
-.002604\,\re^9\right). 
\end{eqnarray}
For Reynolds number~$\re>11$ the highest powers in~$\re$ dominate.
Observe that for these Reynolds numbers both curvature and torsion
may easily reverse the sign of of the skewness paramater~$E$ through 
the combination
\begin{displaymath}
    +\re^7\left[.001067\,\kappa^2 +.003968\,\re^2(\kappa^2\tau^2
    +{\kappa'}^2)\right]\,.
\end{displaymath}
Again this effect promotes upstream tails in the dispersion.  Although
the terms in $\kappa\kappa'$ and $(\kappa\kappa')'$ may keep
$E$~negative, we prefer to interpret these as representing upstream
memory.

\paragraph{For the dispersion of heat in air} with a Prandtl number of 
$\sigma=0.71$ at~$15^\circ\mbox{C}$ and recalling that $\rr=\re/10$, 
the dispersion coefficient~(\ref{eq:skew}) reduces to 
\begin{eqnarray}
	&&E  =  -0.1242\,\rr^3
		\nonumber\\&&{}
+\kappa^2\left(
-2.884\,\rr
+3.124\,\rr^3
-0.6995\,\rr^5
+0.1645\,\rr^7\right)
		\nonumber\\&&{}
+\kappa\kappa'\left(0.5833
+6.217\,\rr^2
-9.592\,\rr^4
+3.537\,\rr^6
-0.7762\,\rr^8\right)
		\nonumber\\&&{}
+(\kappa^2\tau^2+{\kappa'}^2)\left(2.221\,\rr
+16.93\,\rr^3
-25.07\,\rr^5
+8.940\,\rr^7
-0.3844\,\rr^9\right)
		\nonumber\\&&{}
+(\kappa\kappa')'\left(
-2.782\,\rr
-12.99\,\rr^3
+22.94\,\rr^5
-9.478\,\rr^7
+1.287\,\rr^9\right)
\,.
\end{eqnarray}
For Reynolds number~$\re>40$ the highest powers in~$\re$ dominate.
Observe that for such larger Reynolds number the sign of the skewness
coefficient changes sign sensitively depending upon the torsion~$\tau$,
curvature~$\kappa$, and its gradients.

\subsection{Higher order curvature affects the dispersion}
\label{SShigh}

Computer algebra also straightforwardly determines even higher order
corrections to the dispersion coefficient.  These terms may be used,
for example, to give estimates of the errors in the earlier
approximations.  However, the algebraic expressions quickly become
extremely complicated.  We just extend the analysis to the next order
in curvature, but no higher order in gradients, to obtain the
following correction to the dispersion coefficient~(\ref{eq:deff}):
\begin{eqnarray}
	D&=&  \cdots + \rat{1}{8}\kappa^4 \nonumber\\&&
	+\frac{\pe^2\kappa^4}{48}\left[
\re^{8} \left(\rat{6959456407}{3094629863915520000} \sigma^{4}
% \right.\right.\nonumber\\&&\qquad\left.\left.
+\rat{148720297230839}{464658674066915328000000} \sigma^{2}
\right.\right.\nonumber\\&&\qquad\left.\left.
-\rat{11319036743801}{14297189971289702400000}\right)
\right.\nonumber\\&&\quad\left.
+\re^{6} \left(\rat{21839753491553}{12654103324262400000} \sigma^{2}
+\rat{1800408289399}{2711593569484800000}\right)
\right.\nonumber\\&&\quad\left.
+\re^{4} \left(-\rat{24648813997}{64377815040000} \sigma^{2}
+\rat{5096950451}{21459271680000}\right)
\right.\nonumber\\&&\quad\left.
-\rat{4685593}{348364800} \re^{2}
+\rat{13829}{9216}
	\right]+\Ord{\kappa^6,\delta}\,;
	\label{eq:dif4}
\end{eqnarray}
or approximately
\begin{eqnarray}
	D &\approx&  \cdots + \rat{1}{8}\kappa^4 \nonumber\\&&{}
	+\frac{\pe^2\kappa^4}{48}\left[
\left(\frac\re{10}\right)^{8} \left(0.22 \sigma^{4}+0.032 \sigma^{2}-
0.079\right)
\right.\nonumber\\&&\quad\left.{}
+\left(\frac\re{10}\right)^{6} \left(1.7 \sigma^{2}+
0.66\right)
+\left(\frac\re{10}\right)^{4} \left(-3.8 \sigma^{2}+
2.4\right)
\right.\nonumber\\&&\quad\left.{}
-1.3 \left(\frac\re{10}\right)^{2}
+1.5
	\right]+\Ord{\kappa^6,\delta}\,.
	\label{eq:dif4n}
\end{eqnarray}
I also computed the dispersion coefficient to the next correction, with
errors~$\Ord{\kappa^8,\delta}$.  Then recalling the Dean
number~$\dn=2\sqrt\kappa\re$, the dominant terms for the coefficient of
dispersion of material in liquids, large Schmidt number~$\sigma$, are:
\begin{eqnarray*}
	D&\approx&\frac{\pe^2}{48}\left[ 1 
- 0.3754\left(\frac{\dn^2\sigma}{400}\right)^2  
+ 0.2249\left(\frac{\dn^2\sigma}{400}\right)^4
- 0.1388\left(\frac{\dn^2\sigma}{400}\right)^6
\right.\\&&\left.{}\vphantom{\frac{\dn^2\sigma}{400}}
+\Ord{\dn^{16}\sigma^8}\right]\,.
\end{eqnarray*}
This expression is valid for small enough $\dn^2\sigma$\,.  Using the
additional information \cite[\S4.3]{Johnson86} that the limit at
large~$\dn^2\sigma$ is approximately~$0.20$\,, I construct the
following Pad\'e approximant in terms of $\alpha={\dn^2\sigma}/{400}$
\begin{figure}[tbp]
	\centering
	\includegraphics[width=0.8\textwidth]{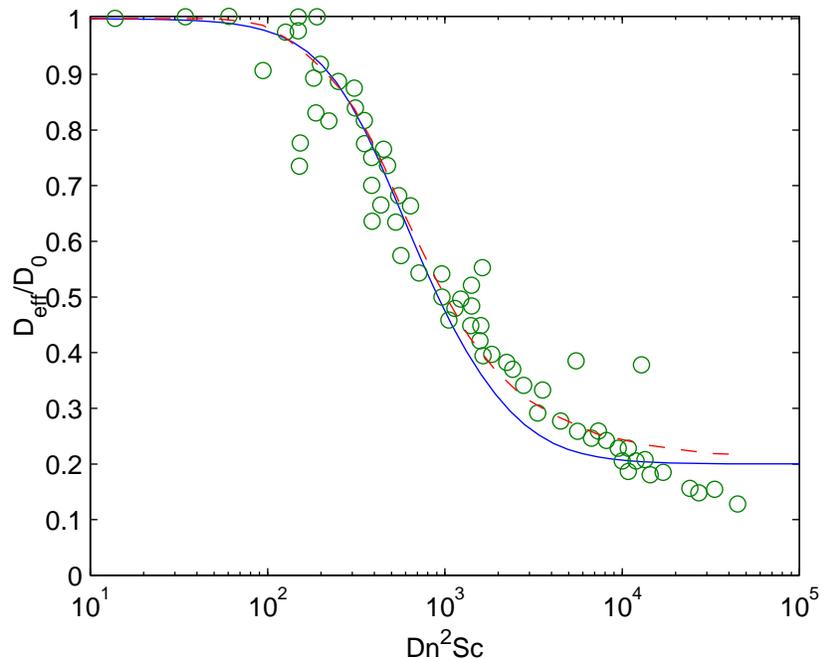}
	\caption{comparison of the Pad\'e approximant~(\ref{eq:deffrat})
	(solid) with experimental estimates (circles) collated by
	Johnson~\protect\cite{Johnson86}, Fig.~9, and the predictions
	(dashed line) of their spectral method based on Poiseuille flow.}
	\label{fig:deffrat}
\end{figure}%
\begin{equation}
	D\approx\frac{\pe^2}{48}\times
	\frac{1+0.3068\,\alpha^2+0.007811\,\alpha^4}%
	     {1+0.6822\,\alpha^2+0.03905\,\alpha^4}\,.
	\label{eq:deffrat}
\end{equation}
See in Figure~\ref{fig:deffrat} that this expression for the dispersion
coefficient matches reasonably well with experiments over the whole
range of $\dn^2\sigma$\,.

% Torsion enhances dispersion for medium Reynolds numbers so do not
% expect it to cause the lowering of~$D$ seen experimentally ---
% possibly the discrepancy is due to the variance deficit affecting
% results in pipes that are too short for such high Peclet number flows.

\section{Conclusion}

Computer algebra handles the considerable details of deriving the
complicated expressions describing dispersion in generally curving
pipes, \S\ref{Sca}.  Fixing one of the fluid flux or the mean pressure
gradient affects the dispersion, \S\ref{sec:intro}, and throughout I
present results for the appropriate case of fixed fluid flux.  The
Pad\'e approximation~(\ref{eq:deffrat}) for the dispersion in a
constant curvature pipe is reasonably accurate over the entire range of
Dean numbers.  When the pipe's curvature and torsion vary, much of the
effects of variations upon the dispersion may be recast as an upstream
memory.  Overall, torsion~$\tau$ in the pipe seems to have little
effect on the dynamics except for a sensitivity, in combination with
the curvature~$\kappa$, of the skewness, \S\ref{sec:skew}.  The
skewness coefficient is very sensitive to curvature, hence is easily
made positive, and which may thus explain the observations of long
upstream tails in the dispersion of material in rivers.

\appendix
\section{Computer algebra derivation}
\label{Sca}

Computer algebra is a very powerful means to derive asymptotic
expansions.  In order for other people to reproduce and verify the
results recorded herein, I list here the core of the program used to
derive the asymptotic expansions; obtain the full program by request.

The computer algebra program was written in  
\red\footnote{At the time of writing, information about {\red} was 
available from Anthony C.~Hearn, RAND, Santa Monica, CA~90407-2138, 
USA. \url{mailto:reduce@rand.org} There were demonstration versions of 
\red\ freely available at \url{ftp://ftp.zib.de/pub/reduce/demo} or 
\url{ftp://ftp.maths.bath.ac.uk/pub/algebra}\,.} to calculate the 
asymptotic expansions of the centre manifold models described in this 
article.

There is a lot of detail to the computer algebra program.  However, the
key to the correctness of the results is the coding of the governing
equations which forms the key part of the core printed here.  The
algorithm iteratively drives to zero the residuals of these equations,
see Roberts~\cite{Roberts96a} for a generic description of the algorithm.
Thus the details about how the residuals are reduced are not vital,
only that they are correctly computed and ultimately zero.

{\footnotesize
\verbatimlisting{pipedcore.red}
}

Observe that the pressure is decomposed into a mean gradient and a
cross-pipe fluctuating component.  There is code to adjust the mean
pressure gradient to ensure a constant mean fluid flux.  To adapt to
the traditional fixed pressure gradient in a helical or toroidal pipe,
one just needs to omit the modifications.  However, the results are
then inappropriate to a pipe with varying curvature or torsion.

\afivepage

\bibliographystyle{plain}\bibliography{ajr,bib,new}

\end{document}